# Subthreshold Jitter in VR Can Induce Visual Discomfort

SAMUEL J. LEVULIS, Reality Labs, Meta, United States of America
KEVIN W. RIO, Reality Labs, Meta, United States of America
PABLO RAMON SORIA, Reality Labs, Meta, Switzerland
JAMES WILMOTT, Reality Labs, Meta, United States of America
CHARLIE S. BURLINGHAM, Reality Labs Research, Meta, United States of America
PHILLIP GUAN, Reality Labs Research, Meta, United States of America

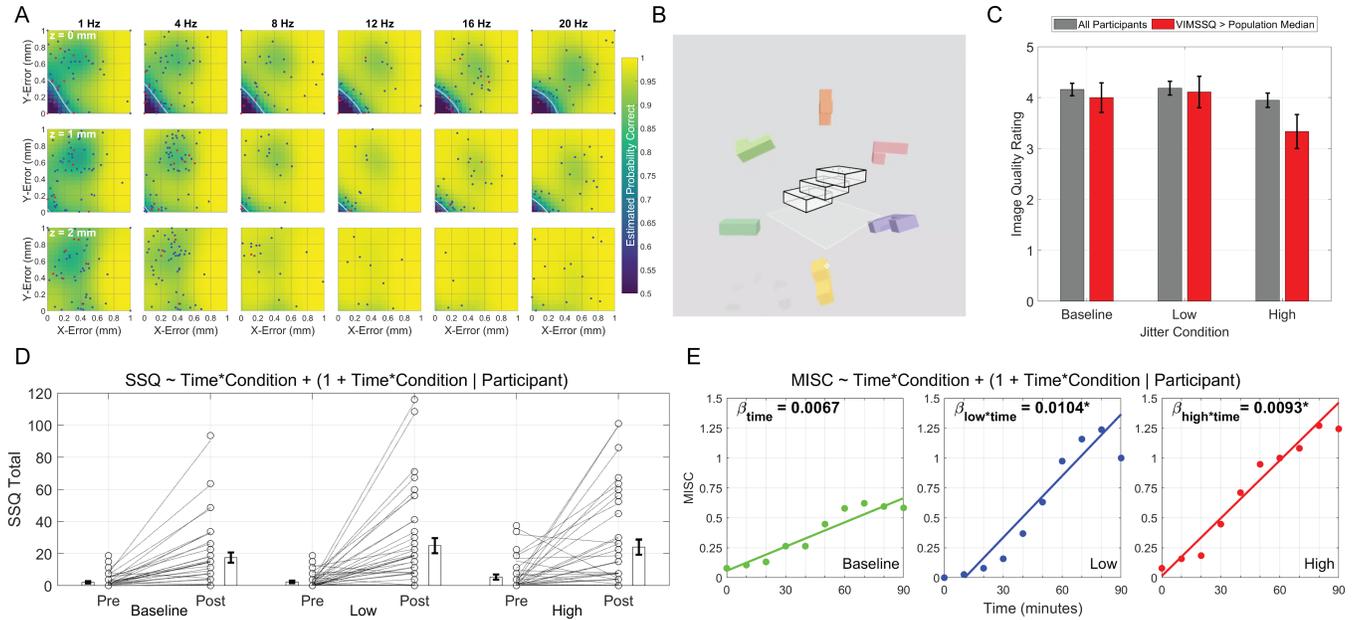

Fig. 1. Virtual reality headmounted displays (VR HMDs) aim to render and present perspective-correct 3D environments that are consistent with a user's movement. A number of hardware and software features in the HMD can introduce discrepancies between geometrically-accurate projection and what a user actually sees. This may include errors in headtracking and eyetracking accuracy, improper HMD fit, incomplete distortion correction, and more. In this work we focus on jitter, a broad class of errors that results in dynamic displacement of the render camera relative to the HMD's center of projection (which ideally coincides with the user's eye position). (A) We first introduce a psychophysical experiment spanning four dimensions of possible jitter error (x, y, and z and temporal frequency) to identify jitter detection thresholds. (B) We next run a repeated-measures comfort study to examine the potential impacts of subperceptible levels of jitter on visual comfort. This study consists of three, 90-minute gameplay sessions of the VR title *Cubism* to investigate visually induced motion sickness (VIMS) in a naturalistic context across three jitter conditions. (C) While subjects overwhelming do not qualitatively comment on jitter artifacts, there is a statistically significant reduction in image quality ratings in the high jitter condition relative to baseline (3.95 vs. 4.16 average rating). (D) Linear mixed effect modeling does not identify statistically significant differences in VIMS symptomology as measured by pre- and post-condition Simulator Sickness Questionnaire score. (E) Linear modeling with the Motion Illness Symptoms Classification (MISC) survey does identify statistically significant differences in comfort across both jitter condition and individual susceptibility (not shown here).

Visual-vestibular conflicts (VVCs) are a primary contributor to visually induced motion sickness (VIMS) in head-mounted displays (HMDs). However, virtual reality (VR) comfort studies often rely on exposing seated or standing users to experiences with high intensity visual motion (such as roller coasters). These drastic VVCs tend to induce pronounced VIMS symptoms that can be reliably detected across individuals using common survey measures. The conclusions from studies using these extreme motion-based conflicts may not accurately generalize to naturalistic use cases in VR where efforts are made to minimize, rather than maximize, VIMS symptoms. In this work, we show that a subthreshold visual-vestibular conflict can induce measurable discomfort during naturalistic, long duration use. We first present a psychophysical study, conducted outside of an HMD, to rigorously identify the perceptual thresholds for sinusoidal noise in render pose (i.e., jitter) resulting in erroneous 3D motion of rendered content. We next introduce subthreshold levels of jitter to a Meta Quest 3 VR HMD and demonstrate that this can induce visual discomfort in participants playing the commercially-available game Cubism across a three-session, repeated-measures study. Importantly, we did not identify statistically significant comfort differences between control and jitter conditions with traditional pre- and post-test comparison of Simulator Sickness Questionnaire (SSQ) scores. Significant differences were only identified using the Motion Illness Symptoms Classification (MISC) survey administered every 10 minutes across each 90 minute session. This highlights the benefits of incorporating time-resolved data points and suggests that lightweight, more frequent surveys may be important tools for measuring visual discomfort in more ecologically-valid scenarios.

## 1 INTRODUCTION

One of the most critical and complex problems to address in virtual reality (VR) head-mounted displays (HMDs) is mitigating visually-induced motion sickness (VIMS) and visual discomfort. Correspondingly, a significant body of work has emerged to identify important contributing factors and correlates to visual discomfort that arise from visual-vestibular conflicts (VVCs) [2, 11, 14, 51]. Despite decades of research, there are no validated frameworks or models that provide calibrated estimates of the relationship between HMD components, demographics, time and VVC-based discomfort [64].

The Simulator Sickness Questionnaire (SSQ) remains the gold standard to measure VIMS symptoms [37], but self-report surveys suffer from low-sensitivity and high variance, so VIMS studies often rely on extremely large VVC to induce large amounts of discomfort (e.g., using roller coaster experiences to show moving content to stationary observers) [15, 19, 23, 24, 60]. Unfortunately, these studies may have limited applicability to realistic scenarios where experiences and hardware are designed to maximize, rather than minimize, visual comfort. Even with hardware and software designed to be comfortable, VIMS remains a problem that afflicts many HMD users due to multiple potential sources of VVC. Many sources of VVC that can induce discomfort and/or perceptual artifacts result in geometric distortions that add error to the stereoimages seen by users in HMDs including pupil swim [22], tracking errors [54], and latency [9, 54, 57]. However, even second-order artifacts unrelated to perspective projection geometry such as the vergence-accommodation conflict have been shown to induce discomfort [30, 61].

Given the numerous potential sources contributing to VVC in HMDs, the significant effort required to conduct rigorous in-lab evaluations of VIMS (even for just a single potential source of VVC), and the lack of first-principles models that can predict the impacts of a particular VVC, identifying and ranking the relative significance of various conflicts is infeasible. One scalable alternative to measuring the VIMS impact of a specific VVC contributor is to use psychophysical paradigms to identify detection thresholds [25]. Psychophysical thresholds are more reliable and consistent across participants compared to qualitative measures and reducing a visual artifact to a subthreshold value can preserve the immediate perceived visual quality and user experience. However, if the VVC in question is merely undetectable, but not entirely eliminated, an important question remains unanswered: Can subthreshold levels of VVC induce discomfort?

This question is particularly relevant in the context of systems-level HMD design where one contributor to VVC can often be improved at the expense of another. For example, in a visual inertial odometry tracking system, data from an HMD's inertial measurement unit (IMU) may be filtered to reduce noise (i.e., jitter) at the potential cost of increased tracking latency [49]. Similarly, eye-tracking data used to update user pupil position for ocular-parallax rendering [59, 71] could be filtered to avoid jitter at the expense of increased latency. These are important tradeoffs because discrepancies between render camera location, viewing position, and the centers of projections of the HMD displays can introduce errors in world-locked rendering [26] and induce visual discomfort [32]. In this work, we demonstrate that sinusoidal jitter error added to an HMD's IMU which does not result in perceptible image quality artifacts can lead to elevated discomfort during naturalistic HMD use through the following contributions:

- We present a psychophysical jitter detection experiment in a jitter-free stereoscopic display and identify peak jitter sensitivity of 0.77 arcmin at 12 Hz. These results are used to design an in-HMD comfort study on subthreshold jitter.
- We introduce sub-threshold levels of jitter in a consumer VR HMD and evaluate visual comfort in the game Cubism. This provides a naturalistic and engaging experience for our three session, repeated measures study (270 total minutes of HMD use). We show that sub-threshold jitter can introduce VIMS and further identify the time course of visual discomfort over different levels of jitter.
- We identify discrepancies between measures of visual discomfort using the Motion Illness Symptoms Classification (MISC) and SSQ surveys and highlight a case where repeated administration of MISC is more sensitive to visual discomfort. We additionally show with repeated-MISC measurements that a person's susceptibility to VIMS is correlated to more adverse discomfort and eye strain, but only at higher levels of jitter.

## 2 RELATED WORK

*VIMS in HMDs.* Many components of an HMD are thought to exaggerate VVCs and thus lead to increased VIMS, including hardware features such as frame rate [70], lens distortions [22, 66], and field of view [11], as well as software features such as motion to photon latency [18, 54, 57], and virtual locomotion profile [8, 12, 12, 13, 20, 42, 69]. The impact of these features on a user's comfort is further moderated by demographic characteristics, e.g., age, gender and prior susceptibility to experience VIMS [3, 4, 21, 56]. Current commercially available VR HMDs have made significant progress in addressing hardware and software impacts on VIMS relative to early implementations [35], yet even in modern devices, subsets of potential users are known to experience VIMS in part due to fit considerations (e.g., insufficient coverage of interpupillary distance supported by available lens interaxial distance settings) [63].

Within this expansive body of work, few studies consider visual discomfort in naturalistic longer-duration use cases and contexts [11, 60]. A meta-analysis of 55 studies by Saredakis et al. [60] found that 91% of VIMS studies were less than 20 minutes of total exposure (sometimes involving multiple conditions) and often involved virtual roller coaster-like experiences where stationary observers were moving in the virtual environment. It is unclear if conclusions about the impact of an HMD feature on VIMS based on exposure to intentionally exaggerated VVCs or short exposure durations directly apply to naturalistic longer-duration HMD use [36, 65]. It is especially challenging to asses the relative importance of different sources of VVC because the field has yet to build a general and predictive theoretical framework for comfort due to the complex underlying nature of VIMS [10, 11, 64].

A significant barrier to building such a framework is simply conducting VIMS studies that generate durable quantitative results. VIMS is highly dependent on inter-individual differences and exposure time [11, 60, 64], so the "highest signal" research design one can employ includes a fully repeated-measures implementation (i.e., all participants complete all conditions on different days) with measurements of VIMS symptom severity taken at regular intervals across a session [7, 33]. Studies employing these methods are likely to observe subtle differences in VIMS that may not be identified when using alternative methods, especially when paired with long duration sessions to increase exposure to VVC and, in principle, the corresponding VIMS response. For example, a recent investigation found comfort impacts from increased lens distortion after participants played a VR game for 80 minutes, but observed no measurable effect for participants who played only 30 minutes under equivalent testing conditions [66]. While the benefits of long duration exposure [48, 50, 65] and repeated-measures designs [28, 47] in detecting VIMS (and differences across conditions) are clear, studies that leverage a combination of these techniques are rare, presumably due to the immense effort and logistical challenges associated with running every participant for multiple 60+ minute sessions.

*VIMS Measurement Methods and Study Design.* One reason naturalistic comfort studies must be so long is because VIMS measurement methods are noisy. While a key focus in VIMS research is developing objective measures of user behavior that can offer more accurate and reliable estimates than surveys [17, 40, 58, 64, 66], survey-based self-reporting of discomfort has historically been the gold-standard for visual comfort evaluation [29, 36, 60]. The most widely adopted survey for VIMS assessment is the SSQ survey [37], which interrogates 16 symptoms that are combined to yield a "total" value based upon three subscales. The SSQ survey has been used across decades of studies, and this backlog of work allows for inter-study comparisons via meta-analyses [60]. However, the time required to complete a 16-item survey prevents the SSQ from being an optimal device for measuring VIMS during headset use (i.e., while in an HMD), positioning it best for use before and after a session. Shorter surveys such as the single item Motion Illness Symptoms Classification survey (MISC, often referred to as the "Misery Scale") [7, 68] and Fast Motion Sickness (FMS) survey [38] rectify this constraint and afford use during an experimental session, but suffer from reduced resolution of symptomology. Thus, there is no "silver bullet" survey that is uniquely best to use for all purposes. Here, we pair the MISC and SSQ surveys in a unified measurement protocol to maximize the potential sensitivity of our survey analyses to measure VIMS differences as a function of jitter.

*Jitter and World-Locked Rendering.* HMD render and viewing geometry assumes that the render cameras, center of projection (CoP) of the HMD displays, and the user's eyes are all co-located in order to render, present, and perceive accurate perspective 3D geometry [62]. Jitter is a dynamic error that displaces the render camera away from the HMD CoPs and users' eyes [32]. Similar to latency, jitter introduces erroneous motion into the rendered percept, degrading world-locked rendering (WLR) accuracy. The perceptibility of WLR errors have been examined in AR and VR [26, 34, 45, 72], but few studies have examined human sensitivity to inaccurate world motion as a function of temporal frequency. Tyler and Tores [67] measured psychophysical thresholds for jitter from 0.1 Hz to 30 Hz and report peak sensitivity between 5-20 Hz for two individuals. Comfort studies have also been conducted to relate jitter frequency to VIMS. Palmissano et. al [55] found that jitter produced greater changes in discomfort with increasing frequency when comparing 0.8, 1.8, and 7.4 Hz. Dennison et. al [16] found significant increases in discomfort when users were exposed to jitter-like head pose perturbations lasting 260 ms duration at 2 Hz. These latter two studies introduced relatively large, suprathreshold jitter perturbations, and measured discomfort over short durations (<10 min). The current study is the first, to our knowledge, that investigates whether small, subthreshold jitter can accumulate to exacerbate VIMS over long durations (90 mins) which has important implications for optimizing the design of tracking, rendering, and display HMD subsytems to support comfortable use over extended sessions.

## 3 EXPERIMENT 1

We first use psychophysical methods to establish a high-confidence measure of human sensitivity to jitter on a world-fixed stereo 3D display (i.e., the jitter added to the stimulus is completely characterized and systematically manipulated by our experimental protocol). This value is used to establish the limits for sub-threshold jitter that are used in our in-HMD study in Experiment 2.

### 3.1 Apparatus

Participants were seated one meter away from a 97" LG OLED TV (OLED97G2PUA), resulting in a 114°H x 82°V field of view. This produced an effective resolution of 34 pixels per degree. A pair of active shutter glasses were synchronized to the display with a photodiode to enable temporally interlaced stereoscopic presentation at 120 Hz (60 Hz per eye). Perspective-correct binocular images for the left and right eyes were rendered using each participant's interpupillary distance (IPD) as measured with a pupilometer, and binocular rendering was done asynchronously to mitigate potential depth artifacts from asynchronous stereo viewing [31].

### 3.2 Stimuli

Participants viewed a paragraph of Lorem Ipsum text spanning 44.10°H x 44.10°V field of view. Black text was rendered on a white background, using the font Optimistic, with a .60 deg (10.47 mm) lowercase x-height. These parameters resulted in 27 lines of text. The text sizes used are broadly consistent with standard recommendations for reading text in print [43] and VR [52].

To add jitter, the user's head pose was (incorrectly) displaced in 3D positional coordinates, which resulted in a perspective-*incorrect* view of the virtual scene. Jitter was constantly applied during stimulus presentation with a sinusoidal time course whose frequency varied between 1-20 Hz. Jitter amplitude was varied along all three positional axes (X, Y, Z; corresponding to horizontal and vertical dimensions in the 2D display plane, and depth, respectively) independently. These values were chosen to align with the spatiotemporal range over which humans are maximally sensitive to jitter [1]. Thus, the stimulus parameter space consisted of four dimensions: the jitter error amplitude in X, Y, and Z spatial dimensions, plus the frequency at which those errors were applied.

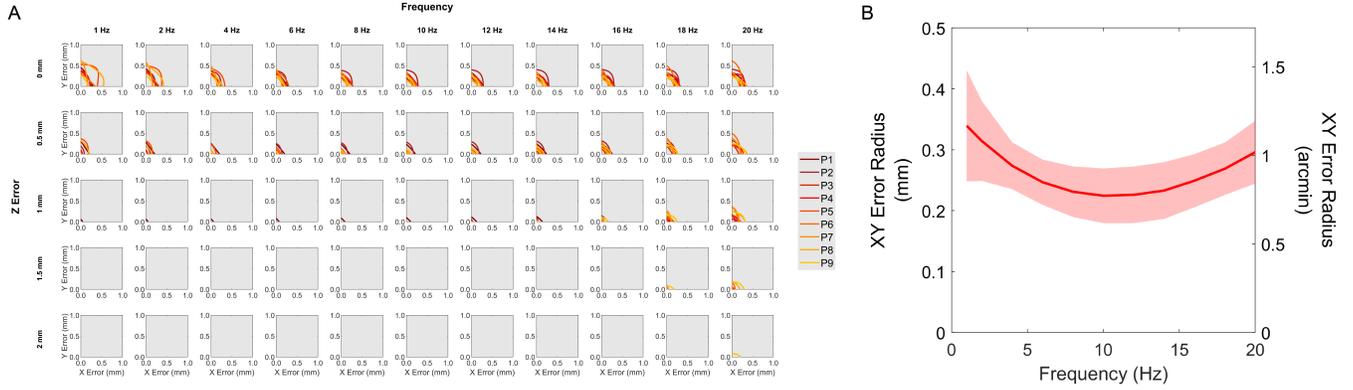

Fig. 2. Psychophysical data on jitter perception. (A) 75% detection threshold contours for each participant, derived from a Gaussian process model fit to response data on a two-interval forced choice task. All participants are plotted in each subplot, four stimulus dimensions are represented: Jitter error amplitude in X and Y (X/Y axes of each subplot); jitter error amplitude in Z (rows); frequency of jitter errors (columns). (B) Summary of 75% detection radii (derived from detection contours) across all participants. Solid line represents mean; shaded region represents 95% confidence interval.

### 3.3 Procedure

Each experimental session lasted 90-120 min. On each trial participants viewed two text samples consecutively. Viewing duration per trial interval was controlled by the participant, with a mean viewing duration of (approximately) two sec. Positional jitter was added to one interval, with order randomized on each trial. Participants completed a two-interval forced-choice (2IFC) task after each trial, and were asked to report which presentation of text appeared more stable. A unique combination of jitter error amplitudes (X, Y, and Z) and frequency was generated and presented on each trial using the Adaptive Experimentation in Psychophysics (AEPsych) framework [53] for N-dimensional stimulus parameter spaces. Participants completed between 811-1000 trials, depending on task completion speed.

### 3.4 Participants

Nine adults completed the experiment. All participants were tested for normal or corrected-to-normal visual acuity and stereoacuity. Study protocols were IRB approved.

### 3.5 Analysis

Each participant's correct and incorrect responses were used to generate an estimate of the psychometric function across the 4D parameter space, using the Gaussian process model described in [44, 53]. Using this model, a contour representing 75% probability of correct response can be extracted along arbitrary dimensions in stimulus parameter space; these contours represent a boundary between predominantly correct and predominantly incorrect responses, and serve as an estimate of the detection threshold. One participant's example data is shown in Figure 1A where blue and red dots represent correct and incorrect trials across combinations of x, y, and z jitter and frequency. The 4D stimulus parameter space was divided into a 2D array of 2D spaces, as shown in Figure 2A. Each subplot represents X and Y error amplitude (along the subplot's x- and y-axes, respectively), with rows corresponding to Z error magnitude and columns corresponding to frequency.

The 75% detection threshold contours from these slices of parameter space were extracted and further analyzed. Two dimensional reductions were made to facilitate interpretation. First, subsequent analyses were limited to trials with negligible variation in Z (0 < Z error amplitude < 0.5 mm), as the effect of Z errors was observed to be significantly smaller than those of X and Y errors. Second, the X and Y error dimensions were collapsed by computing the radius of each detection contour in 2D (X/Y error amplitude) space. These detection contour radii provide a single XY jitter error amplitude value that represents perceptual sensitivity, which can then be plotted against the remaining dimension of frequency. Figure 2B shows the mean XY error radius computed across all participants as a function of frequency.

### 3.6 Results

Across participants ($N$=9), sensitivity to jitter was greater in the XY axes compared to Z (e.g., in Figure 2A almost all XY thresholds are below 0.5 mm when Z = 0 mm in the first row, but some XY jitter is still undetectable when Z = 0.5 mm in the second row). We therefore analyze the data in the XY axes across temporal frequency, setting the value of Z equal to zero. The mean XY error 75% detection contour radius was lowest (indicating the highest perceptual sensitivity) at 12 Hz, with an XY jitter error amplitude (i.e., half peak-to-trough magnitude) of 0.23 mm (95% CI: 0.18-0.27 mm), or 0.77 arcmin in angular units at the viewing distance of 1 m. The mean XY error radius increased for higher and lower frequencies. Data were fit using a second-order polynomial of the form $y = ax^2 + bx + c$, with estimated coefficients $a$=0.0010, $b$=0.024, $c$=0.36, and demonstrated a strong fit ($R^2$=0.99). This suggests that the change in perceptual sensitivity closely follows a quadratic trend as frequencies deviate from the peak sensitivity (lowest XY error radius) at 12 Hz.

## 4 EXPERIMENT 2

Next, we present details on a multi-session study to investigate the effects of jitter on visual comfort. Thirty-eight participants completed our study, with each user spending 270 minutes in HMD over three conditions.

### 4.1 Hardware and Stimulus

Participants wore a Meta Quest 3 HMD (v72.0) outfitted with aftermarket Quest Elite head straps to support more comfortable ergonomic fit over 90 minutes. Participants played Cubism, a virtual 3D puzzle game in "immersive VR mode." Jitter was introduced by superimposing a continuous sinusoidal signal onto real-time inertial measurement unit (IMU) measurements in the axis aligned with gravity. This IMU signal is an input to the headtracking system, which results in errors in computing the head pose used for rendering. Participants were exposed to three different levels of jitter across separate days: no added jitter (baseline), low jitter, and high jitter. The resulting noise at low and high jitter conditions can be approximated as an 18 Hz sinusoidal error in the head pose up-axis with an amplitude of 0.06 mm and 0.15 mm (i.e, 0.12 and 0.30 mm peak to trough magnitudes) for the low and high conditions, respectively. The angular errors resulting from these pose errors are scene dependent, but when considering the same planar content used in Experiment 1 both low and high jitter conditions are below detection thresholds (Figure 2B).

### 4.2 User Study Protocol

Participants completed 90-minute sessions of Cubism in the baseline, low, and high jitter conditions across three separate days (counterbalanced across participants). In the first session participants completed a demographic survey and the Visually Induced Motion Sickness Susceptibility Questionnaire (VIMSSQ; [39, 41]). Each session, prior to donning the HMD, participants completed the SSQ survey. After donning the HMD participants were instructed to open Cubism and progress through the game at their own pace starting from level one. Upon opening Cubism, MISC and eye strain questions were verbally administered, and then again every 10 minutes (participants remained in HMD) until 90 minutes elapsed. After the final MISC and eye strain questions were administered the participant doffed the HMD, completed the post-condition SSQ survey, and provided open-ended subjective feedback on how they were feeling. They also rated the image quality of the experience and were asked to comment on the HMD image quality during their session. All protocols were IRB approved.

### 4.3 Participants

Thirty-eight adults (24 male, 14 female) between 23 and 75 years of age ($M$ = 39.95, $SD$ = 12.61) completed the study. Participants reported an average of 2.48 yrs of experience with VR ($SD$ = 3.67) and playing an average of 1.95 hrs of VR per week ($SD$ = 5.00). Participants' average VIMSSQ score (susceptibility to VIMS) was 15.35 ($SD$ = 17.96); nine individuals were above the estimated population median of 22 [39].

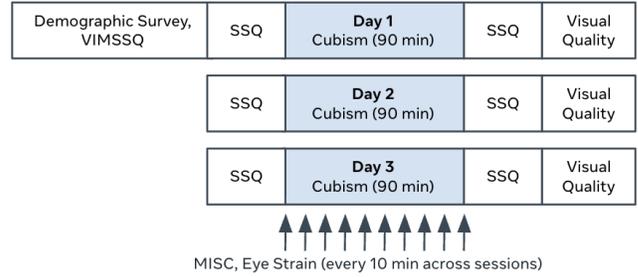

Fig. 3. Experiment 2 study timeline.

### 4.4 Data Analysis

Linear mixed effects models were used to (1) infer the influence of jitter and VIMS susceptibility (VIMSSQ) on VIMS symptoms and eye strain and (2) to predict the progression of symptoms in an average user across time. For each response variable (SSQ, MISC, eye strain), we specified a maximal model by default (i.e., including random intercepts and slopes for all relevant factors) following [5, 6, 27, 46]. This was also supported by the observation of inter-observer variability in VIMS and eye strain symptoms at pre-test as well as in their progression over time (see supplementary materials for model comparison details). Continuous predictors (time, VIMSSQ) were centered prior to analysis. Jitter condition was treated as a categorical variable with baseline as the reference group.

### 4.5 Jitter Perceptibility and Image Quality

At the end of each session, no participants commented about image artifacts that could be attributed to jitter in the baseline or low conditions, and only 2/38 participants noted jitter-related artifacts in the high condition. However, despite the lack of explicit comments about jitter, average image quality ratings across all participants is significantly lower in the high jitter condition (3.95) compared to baseline (4.16), $\beta$ = -0.211, $p$ = 0.032 (Figure 1C).

We explored whether participants noticed jitter (i.e., jitter is above threshold in the high condition), but did not have the vocabulary to effectively describe it; or rated image quality lower due to generally elevated discomfort. There is a theoretical argument that the added jitter in the high condition is above threshold. The content in Cubism is generally much closer than one meter, which means that the added render camera errors will generally induce larger angular errors than the 0.77 arcmins shown in Figure 2 even though the positional errors are below our detection thresholds from Section 3. However, the content in this experiment is experienced naturalistically, and "detection" or noticeability thresholds are likely elevated here compared to a direct A/B comparison of plain text. Conversely, when we looked at the relationship between comfort and image quality ratings, we found a statistically significant correlation between comfort at session completion (both MISC and post-test SSQ total) and image quality ratings. Both factors can plausibly explain, and perhaps both contribute, to why a "subthreshold" artifact leads to lower image quality ratings.

Some evidence also suggests that individuals with higher VIMSSQ scores were more likely to give lower image quality ratings, and that this difference is more pronounced in the high jitter condition (Figure 4). Interestingly, the two individuals who noted jitter-related artifacts are not from the higher susceptibility group. These interactions are only marginally significant and full model output is provided in supplementary materials, we explore other aspects of individual differences in Section 4.7.

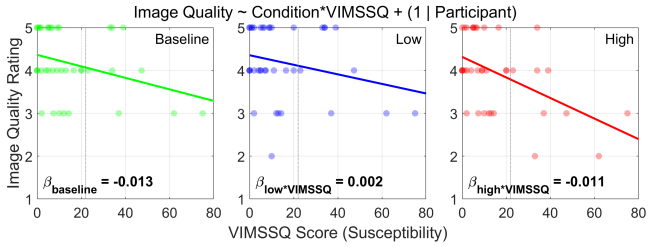

Fig. 4. Image quality ratings are shown for baseline, low, and high jitter conditions as a function of VIMS susceptibility. Across all three conditions there is a marginally significant, negative correlation between VIMSSQ and image quality rating ($p$=0.053). High susceptibility individuals also appear to be more adversely affected by the high jitter condition compared to baseline and low jitter conditions as evidenced by the steeper slope of the linear model output, though this effect is also only marginally significant ($p$=0.056).

### 4.6 Visual Discomfort from Jitter

Based on thresholds identified in Experiment 1 and qualitative responses in Section 4.5 there is strong evidence that jitter was imperceptible in the baseline and low jitter conditions (and weaker evidence for the high jitter condition). We next examine whether a sub-threshold jitter artifact can induce discomfort.

*SSQ Analysis.* We first evaluated the impacts of jitter on reported VIMS symptoms using the traditional comparison of pre- and post-test SSQ Total scores across the three jitter conditions with a General Linear Mixed-Model. Participants' pre- and post-condition SSQ Total scores were modeled as a function of time (pre- vs. post-session) and jitter condition (baseline, low, high). We specified a maximal model accounting for inter-individual differences in pre-test SSQ and inter-individual differences in the effect of jitter magnitude on SSQ over time. Complete output for all models is presented in the Supplementary Material. The main two-way interaction of interest — between jitter condition and time — was not significant for either low or high jitter (Figure 1D).

*MISC Analysis.* We also analyzed how participants' MISC scores changed over time (10 minute intervals from session start to 90 minutes) as a function of jitter condition. Under the maximal model, both low and high magnitudes of jitter caused significantly greater increases in MISC over time relative to baseline (time main effect: $\beta$ = 0.0067, $p$ = 0.003; low jitter * time interaction: $\beta$ = 0.010, $p$ = 0.010; high jitter * time interaction: $\beta$ = 0.009, $p$ = 0.029). This demonstrates that added jitter increased the rate of VIMS symptom onset (Figure 1E).

Indeed, the predicted amount of time the average user could play the game before reaching a MISC score of 1 (cutoff for "uneasiness") was 58 minutes with low jitter, 62 minutes with high jitter, versus over 90 minutes (prediction = 148) in the baseline condition. We used Monte Carlo simulations to estimate 95% confidence intervals on these time-to-symptom-onset predictions (baseline: [84, 390] minutes; low jitter: [36, 120] minutes; high jitter: [37, 145] minutes). To be conservative, for all three predictions, we assumed the same starting MISC score of 0.0526 (empirical average across all groups). Collectively, these findings suggest that subthreshold jitter can significantly accelerate and elevate VIMS symptoms in VR.

*Consistency with Qualitative Feedback.* We observed good general consistency in SSQ and MISC ratings and participants' subjective feedback. All participants who reported MISC ratings indicating Nausea (MISC = 6-8) or Fair to Severe Dizziness or Headache (MISC = 4-5) also made comments directly corroborating this. The MISC and SSQ Total values were also largely in agreement with only two notable discrepancies (P26 high jitter and P45 baseline). These direct comparisons can be seen in Figures 14-16 in the supplement which plots SSQ and MISC responses across individuals.

### 4.7 Effects of Individual Susceptibility

For individuals who reported being more prone to VIMS, jitter had a greater impact on their VIMS and eye strain symptoms across all conditions. Additionally, individuals with higher VIMSSQ scores reported increased VIMS symptoms and eye strain at an even greater rate in the high jitter condition. In other words, elevated VIMS symptoms reported for low jitter in Section 4.6 is proportionately distributed across all individuals based on their underlying susceptibility, but increased discomfort in the high jitter condition disproprtionately impacts higher susceptibility individuals.

*Reduced Model.* Because susceptibility is a per-participant fixed effect, any variance that could be attributed to it in the data would be absorbed by random slopes and interactions in the maximal model, we fit a second model for each response variable with only random intercepts for each participant with VIMSSQ score as a fixed effect. Although this was a worse-fitting reduced model, we only used it for inference about the three-way interaction of susceptibility, time, and jitter condition reported here (see Supplementary Information for model outputs).

*Effect of VIMSSQ.* VIMS susceptibility varied substantially across participants in the study (median VIMSSQ score: 9; mean: 15, median absolute deviation: 7.25; min: 0; max: 75). Among more susceptible participants, the impact of jitter on VIMS symptoms as measured by MISC (but not SSQ) was greater across all conditions (e.g., time * susceptibility interaction: $\beta$ = 0.0003, $p$ = 0.005; also see next section). Similarly, higher VIMSSQ scores were correlated with higher reports of eye strain across all conditions ($\beta$ = 0.010, $p$ = 0.021).

*VIMSSQ Interaction with Jitter Condition.* For high, but not low jitter, more susceptible participants had a significantly faster onset of VIMS symptoms relative to baseline (time * high jitter * susceptibility interaction: $\beta$ = 0.0003, $p$ = 0.019; time * low jitter * susceptibility interaction: $p$ = 0.754). After 90 minutes, mean MISC scores were 3.25 times higher for high vs. baseline jitter in higher susceptibility participants, also crossing 1 (the cutoff for "uneasiness") for high but not baseline jitter. The same interactions in the analogous model for SSQ were not significant. Among more susceptible participants, the impact of jitter on reported eye strain was also greater for high, but not low jitter (although we note that this study cannot discern whether participant responses about eye strain can be decoupled from their general discomfort). In the high jitter condition more susceptible participants' had a significantly faster onset of eye strain symptoms relative to baseline (time * high jitter * susceptibility interaction: $\beta$ = 0.0003, $p$ = 0.0006; time * low jitter * susceptibility interaction: $p$ = 0.164).

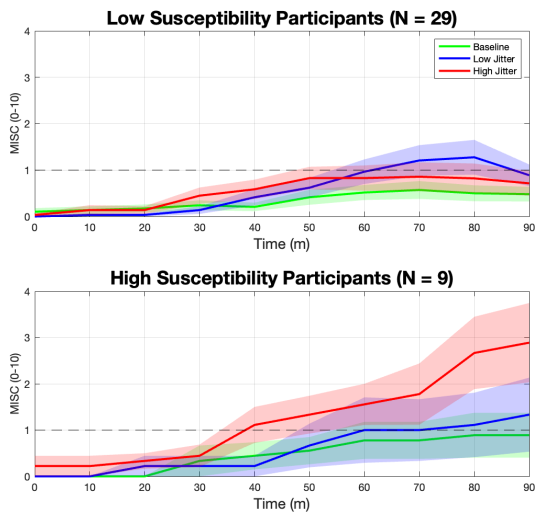

Fig. 5. VIMS symptoms were accelerated and worsened in the presence of high jitter in higher susceptibility participants (N=9). Misery Scale (MISC) progression over time for low and high-susceptibility participants for baseline, low, and high jitter. Curve, mean across participants. Error surface, SEM across participants. Low and high susceptibility categories were defined with respect to population median of VIMSSQ = 22 in this figure (susceptibility was treated as a continuous variable in models).

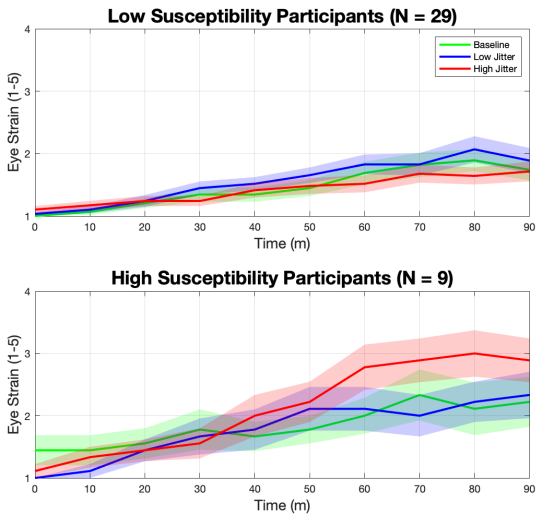

Fig. 6. Eye strain symptom onset occurred earlier and became more severe for higher susceptibility participants in the high jitter condition. Eye strain progression over time for low and high-susceptibility participants for baseline, low, and high jitter. Curve, mean across participants. Error surface, SEM across participants. Low and high susceptibility categories were defined with respect to population median of VIMSSQ = 22 in this figure (susceptibility was treated as continuous variable in models).

## 5 CONCLUSION AND DISCUSSION

We measured device-agnostic perceptual thresholds for jitter detection across temporal frequency and orientation (Experiment 1), and assessed the visual comfort and experience impacts of jitter during long duration (90 minute) naturalistic VR use (Experiment 2). The results of Experiment 1 show individuals are most sensitive to jitter at 12 Hz in VR. Moreover, we found sensitivity to be substantially higher for jitter along X and Y (vertical and horizontal motion) relative to Z (motion in depth). In Experiment 2 we found that subthreshold jitter accelerates and elevates VIMS symptoms for VR users, and that this effect is more pronounced for individuals who are more susceptible to motion sickness. We postulate that jitter produces earlier onset of VIMS due to increases in visual-vestibular conflict. Even if unnoticed, more frequent and higher magnitude errors in render pose relative to user position lead to sensory conflict, which in turn accelerates VIMS.

*Implications of HMD Choice in VIMS Studies.* Differences in HMD hardware are intuitively important when comparing across studies, but our study has shown that small, potentially imperceptible changes in HMD software may also lead to significant changes in VIMS outcomes. Modern HMDs can receive significant software updates over their lifetimes, and even different software versions running on the same hardware may have significant implications when running VIMS studies. This is important to keep in mind when conducting meta-analysis, or when using another study as a comparison or baseline. The safest and most sound way to study the effects of specific VIMS manipulations is therefore to run repeated-measures studies with an unadulterated HMD baseline.

*VIMS Measures over Time.* We found a dissociation in VIMS outcomes as measured by MISC (fast motion sickness question administered every 10 minutes) and SSQ (16-item survey administered pre- and post-session). While the general pattern of SSQ means was consistent with MISC outcomes, no interactions reached statistical significance. Since these data come from the same participants and putatively assessed similar constructs (VIMS), the difference in findings suggests that more frequent, less exhaustive surveys such as MISC may provide a more sensitive measure by better characterizing the time course of visual discomfort in longer duration experiments.

*Individual Susceptibility and Repeated-Measures.* We found that the addition of low jitter increased visual discomfort for all individuals similarly (i.e., there was no significant interaction between VIMSSQ and condition and time). However, we also found that the high jitter condition disproportionately affected higher-sensitivity individuals compared to the rest of our participant population (i.e., there was a significant interaction between VIMSSQ and condition and time). Given the inherent noise in self-report surveys that might affect both VIMSSQ and MISC responses, this interaction between VIMSSQ and the high jitter condition may not have been identified in a between-subjects response. Repeated-measures studies have the potential to identify meaningful and important inter-individual trends that might otherwise be lost. VIMS researchers should consider opportunities for repeated-measures study designs that incorporate lower overhead, but more frequent measures of discomfort like MISC.

## ACKNOWLEDGMENTS

Thanks to Cole Williams, Drew Wright, and Carla Romero for their extraordinary efforts in user study data collection.

# Subthreshold Jitter in VR Can Induce Visual Discomfort
## Supplementary Material

SAMUEL LEVULIS, Reality Labs, Meta, United States of America
KEVIN W. RIO, Reality Labs, Meta, United States of America
PABLO RAMON SORIA, Reality Labs, Meta, Switzerland
JAMES WILMOTT, Reality Labs, Meta, United States of America
CHARLIE S. BURLINGHAM, Reality Labs Research, Meta, United States of America
PHILLIP GUAN, Reality Labs Research, Meta, United States of America

## 1 EXPERIMENT 1

Here we provide the full individual results from the psychophysical experiment. In the figures below white lines represent detection threshold contours for each participant, derived from a Gaussian process model fit to response data on a 2 interval forced choice task. Four stimulus dimensions are represented: jitter error amplitude in X and Y (X/Y axes of each subplot); jitter error amplitude in Z (rows); temporal frequency (columns). Red dots represent trials where the participant incorrectly identified which of the two intervals was rendered with jitter and blue dots represent trials where participants identified the jitter interval correctly. Large values of jitter (upper right within each subplot, and lower rows) generally contain more correct responses as jitter magnitude increases.

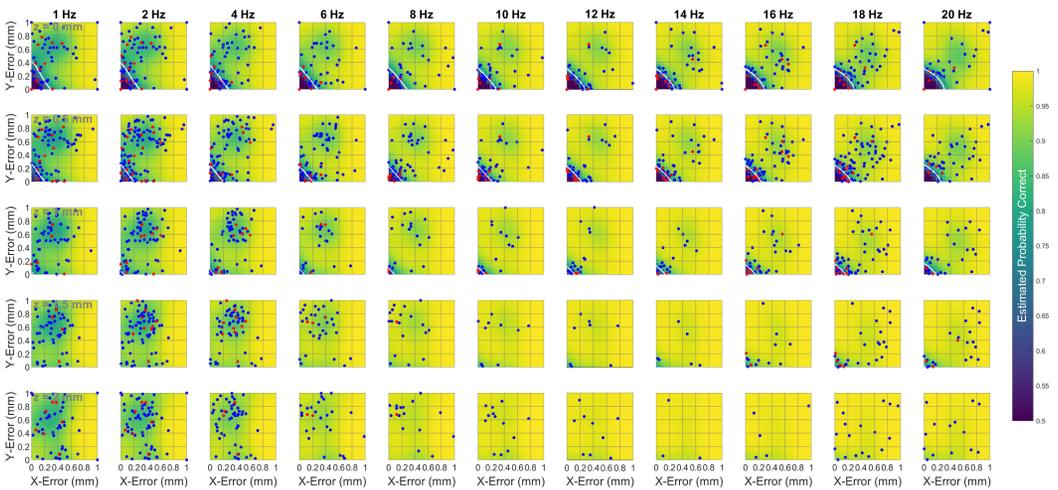

Fig. 1. Participant 1.

2    Samuel Levulis, Kevin W. Rio, Pablo Ramon Soria, James Wilmott, Charlie S. Burlingham, and Phillip Guan

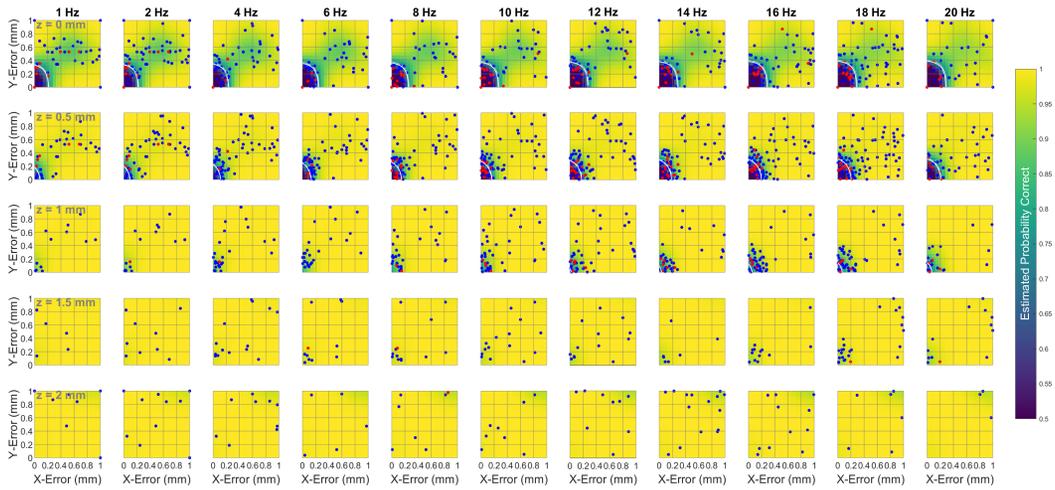

Fig. 2. Participant 2.

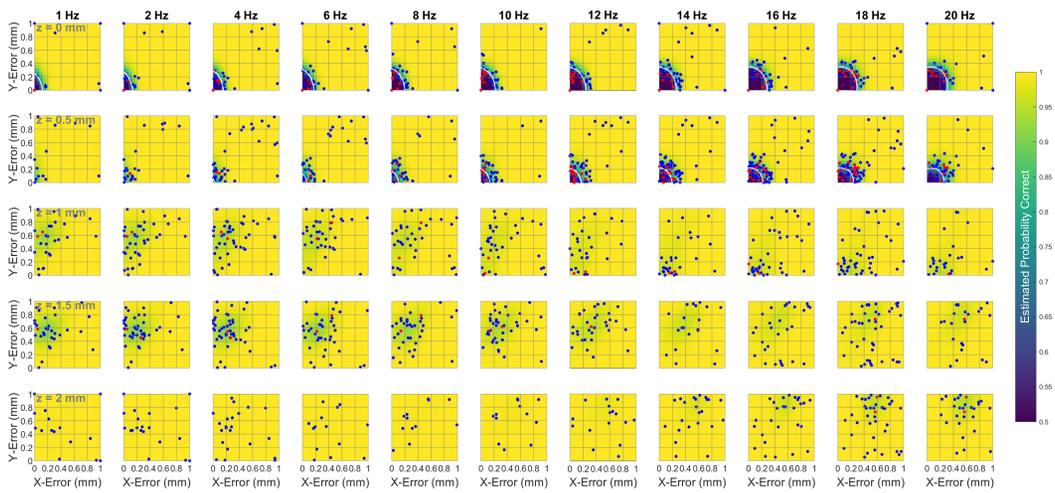

Fig. 3. Participant 3.

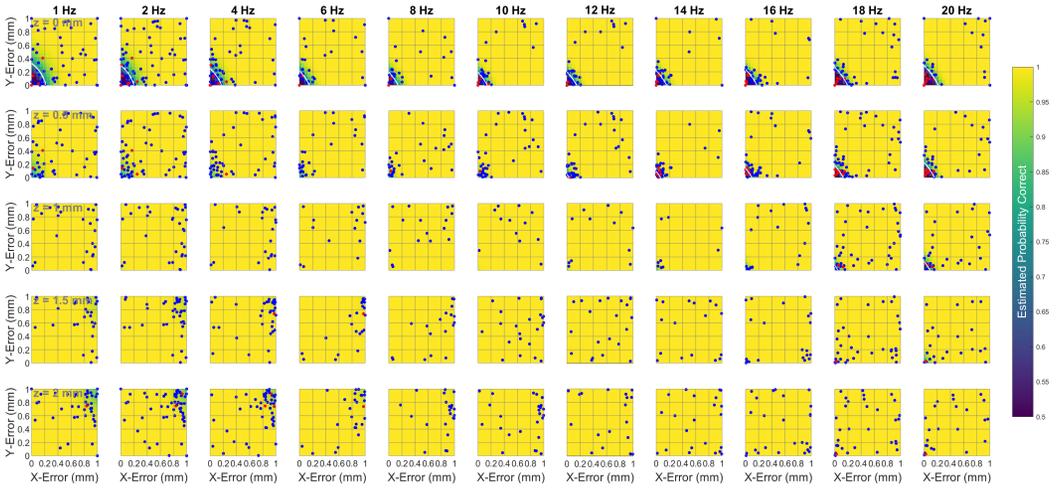

Fig. 4. Participant 4.

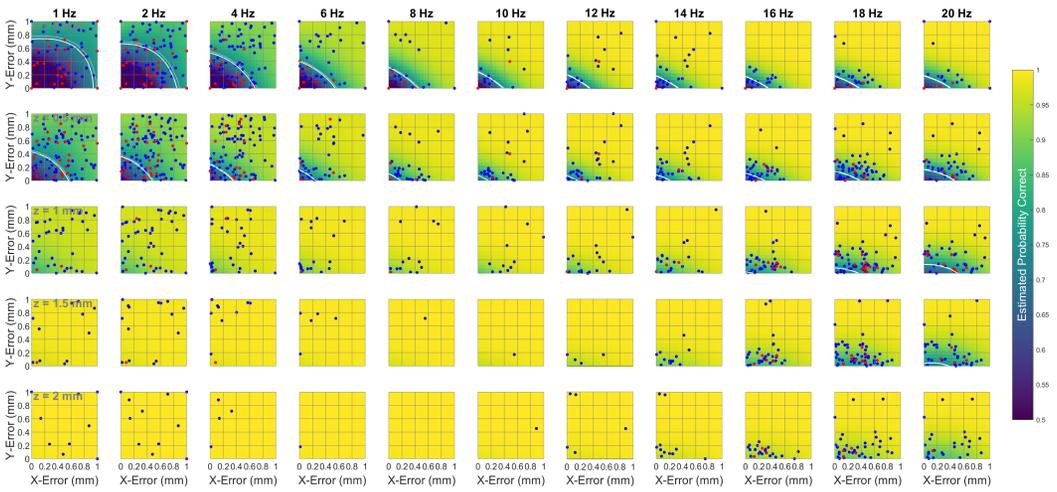

Fig. 5. Participant 5.

4      Samuel Levulis, Kevin W. Rio, Pablo Ramon Soria, James Wilmott, Charlie S. Burlingham, and Phillip Guan

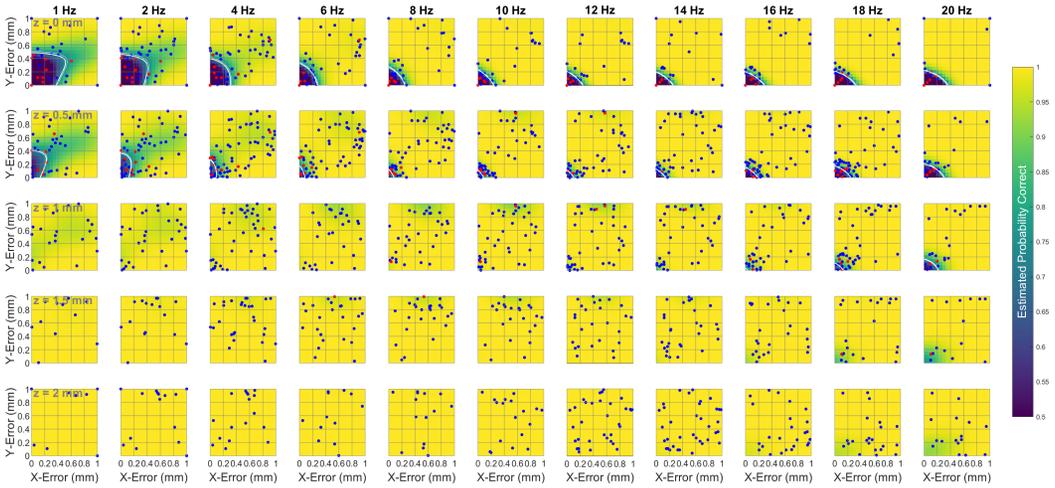

Fig. 6. Participant 6.

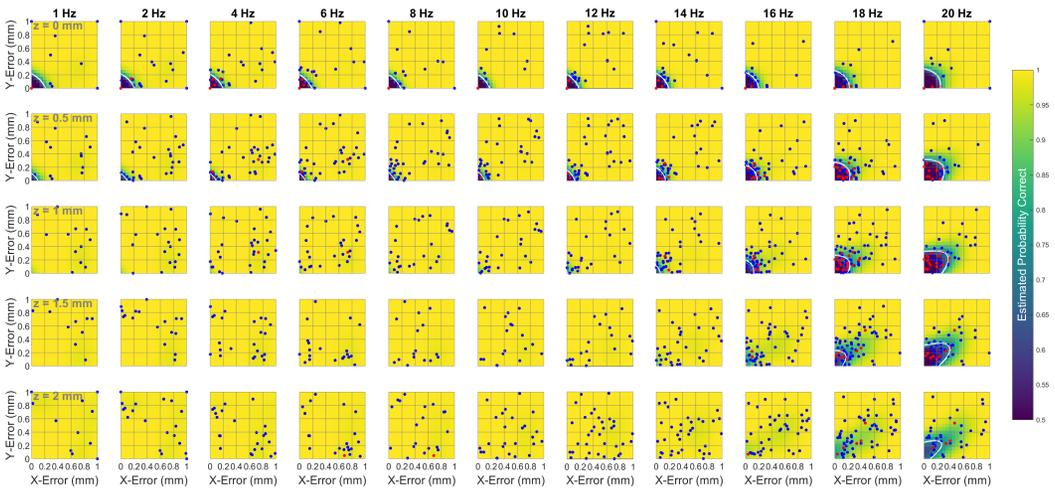

Fig. 7. Participant 7.

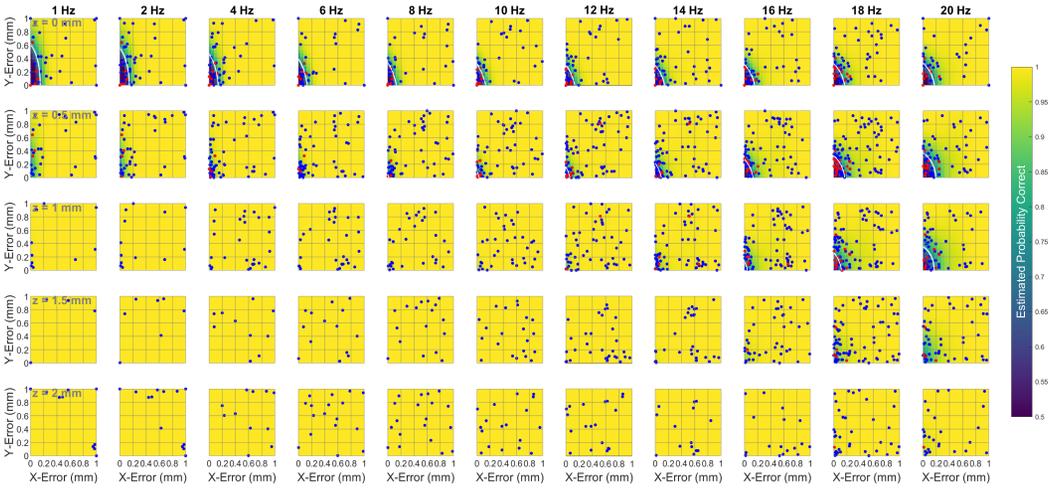

Fig. 8. Participant 8.

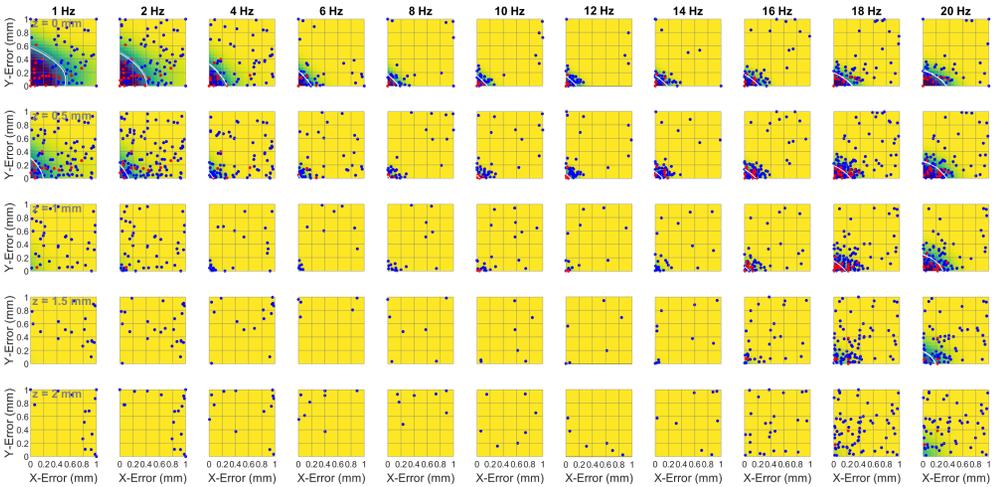

Fig. 9. Participant 9.

## 2 EXPERIMENT 2
### 2.1 Model Selection

We used likelihood ratio tests to perform model comparisons as a complementary way to provide evidence for the inclusion of random slopes and interactions in our models. For each response variable, a reduced model with only random intercepts per participant was significantly worse than the maximal model (e.g., for MISC: likelihood ratio test statistic, 914; $R^2$: 39% vs. 80%). As a second model selection criterion, we compared predictions in real units (time-to-symptom-onset, minutes) under the maximal and reduced model in order to estimate the magnitude of a potential bias in the relevant fixed effect coefficients arising from model mis-specification. The difference in the prediction between models was up to 15 minutes, which we deemed a substantial enough possible error to again support using the maximal model.

## 2.2 Jitter Perceptibly, Image Quality, and VIMSSQ

There was a marginally significant effect of VIMS susceptibility on mean image quality rating as well as a marginally significant interaction between VIMS susceptibility and the high jitter condition (Figure 4). Higher susceptibility individuals generally rated image quality lower in all conditions ($\beta$ = -0.013, $p$ = 0.053). In the high jitter condition, higher susceptibility individuals had even larger image quality rating decreases (relative to baseline) than lower susceptibility individuals ($\beta$ = -0.011, $p$ = 0.056). Here we report the complete results from the random-intercept only mixed-effects linear model (assessing the effects and interactions of jitter condition and VIMSSQ) . For all models we used the MATLAB fitlme package and default maximum likelihood estimation. The model had the form:

$$Image\ Quality = \beta_{Intercept} + \beta_{Low} + \beta_{High} + \beta_{VIMSSQ} + \beta_{Low:VIMSSQ} + \beta_{High:VIMSSQ} + \beta_{Participant}$$

where *ImageQuality* is the reported image quality rating, $\beta_{Intercept}$ is the predicted image quality for the baseline condition (for someone with average VIMSSQ), $\beta_{Low}$ is the predicted change in image quality rating for the low jitter condition (for someone with average VIMSSQ), $\beta_{High}$ is the predicted change in image quality rating for the high jitter condition (for someone with average VIMSSQ), $\beta_{VIMSSQ}$ is the predicted change in image quality rating in the baseline condition for each unit increase in a participant's VIMSSQ, $\beta_{Low:VIMSSQ}$ is the predicted additional change in image quality rating in the low jitter condition (relative to baseline) for each unit increase in a participant's VIMSSQ, $\beta_{High:Time}$ is the predicted additional change in image quality rating in the high jitter condition (relative to baseline) for each unit increase in a participant's VIMSSQ, and $\beta_{Participant}$ is a random-effects parameter that fits individual intercepts to each participant to account for between-participants sources of error. The results of this model are reported in Table 1.

| Parameter | Estimate | $t$ value [DoF] | $p$ value |
|---|---|---|---|
| *Intercept* | 4.158 | 34.165 [108] | 1.013$e$-59 |
| *Low jitter* | 0.026 | 0.271 [108] | 0.787 |
| *High jitter* | -0.211 | -2.169 [108] | 0.032 |
| *VIMSSQ* | -0.013 | -1.954 [108] | 0.053 |
| *Low jitter : VIMSSQ* | 0.002 | 0.413 [108] | 0.681 |
| *High jitter : VIMSSQ* | -0.011 | -1.928 [108] | 0.056 |

Table 1. Linear mixed-effects model results predicting image quality from jitter condition and VIMSSQ. Random intercept only model with baseline condition as the reference group, and VIMSSQ mean-centered.

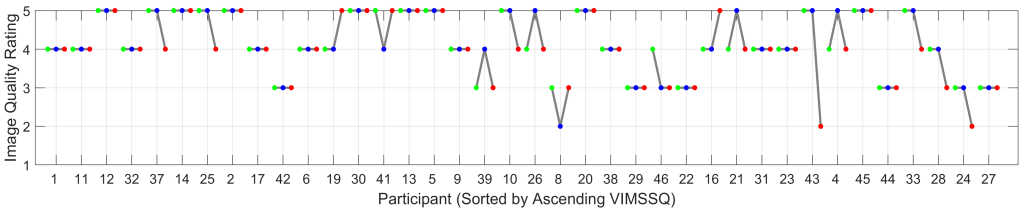

Fig. 10. Individual image quality ratings for each participant in Experiment 2, sorted by VIMSSQ. Green=baseline; blue=low jitter; red=high jitter.

## 2.3 SSQ Models

Here we report the complete results from the maximal linear mixed-effects model (assessing the effects and interactions of jitter condition and time) and random-intercept only model (assessing these same fixed effects but with the addition of VIMSSQ).

In the first (maximal) model assessing only the influence of the combination of jitter condition and time, we fit a mixed-effects linear model (with inclusion of random effects) of the form:

$$SSQ_{Total} = \beta_{Intercept} + \beta_{Low}X_{1ij} + \beta_{High}X_{2ij} + \beta_{Time}X_{3ij} + \beta_{Low:Time}X_{1:3ij} + \beta_{High:Time}X_{2:3ij} + \beta_{Participant}$$

where $SSQ_{Total}$ is the reported SSQ Total value, $\beta_{Intercept}$ is the predicted SSQ Total in the baseline condition at post-test, $\beta_{Low}X_{1ij}$ is the predicted change in SSQ Total at post-test in the low jitter condition (and its random slope to account for between-participants differences in rate of change), $\beta_{High}X_{2ij}$ is the predicted change in SSQ Total at post-test in the high jitter condition (and its random slope), $\beta_{Time}X_{3ij}$ is the predicted change in SSQ Total at pre-test in the baseline condition, $\beta_{Low:Time}X_{1:3ij}$ is the predicted change in SSQ Total at pre-test in the low jitter condition (and its random slope), $\beta_{High:Time}X_{2:3ij}$ is the predicted change in SSQ Total at pre-test in the high jitter condition (and its random slope), and $\beta_{Participant}$ is a random-effects parameter that fits individual intercepts to each participant to account for between-participants sources of error. The results of this model are reported in Table 2.

| Parameter | Estimate | $t$ value [DoF] | $p$ value |
|---|---|---|---|
| *Intercept* | 17.421 | 5.557 [222] | 7.828$e$-8 |
| *Low jitter* | 7.480 | 2.082 [222] | 0.038 |
| *High jitter* | 6.594 | 1.578 [222] | 0.116 |
| *Time* | -15.354 | -5.225 [222] | 4.017$e$-7 |
| *Low jitter : Time* | -7.185 | -1.917 [222] | 0.057 |
| *High jitter : Time* | -3.248 | -0.686 [222] | 0.493 |

Table 2. Linear mixed-effects model results predicting SSQ Total from time and jitter condition. Maximal model with inclusion of random slopes for each effect. Post-test and baseline are the reference conditions.

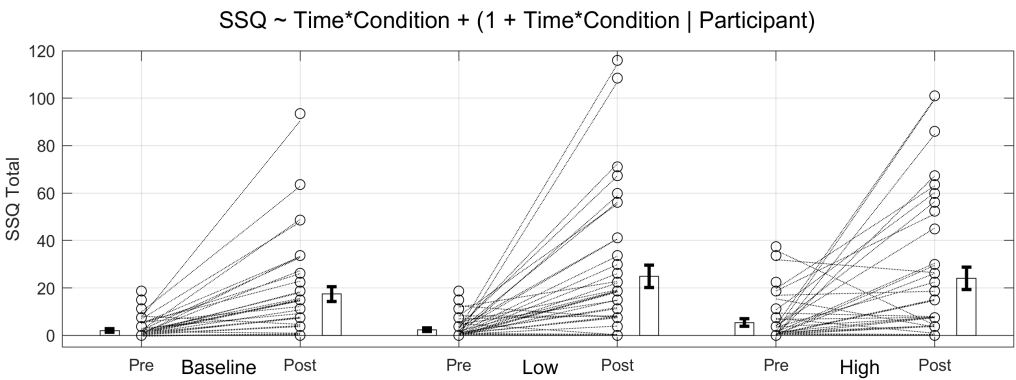

Fig. 11. Linear mixed-effects model results predicting SSQ Total from time and jitter condition. Maximal model with inclusion of random slopes for each effect. White vertical bars depict the mean for each condition (error bars represent +/- 1 SE of the mean.

In the second (random intercept only) model we included the effect of VIMSSQ (which was mean centered) and its interactions. The results of this model are reported in Table 3.

| Parameter | Estimate | $t$ value [DoF] | $p$ value |
|---|---|---|---|
| *Intercept* | 17.421 | 6.101 [216] | 4.816$e$-9 |
| *Low jitter* | 7.480 | 2.117 [216] | 0.035 |
| *High jitter* | 6.594 | 1.867 [216] | 0.063 |
| *Time* | -15.354 | -4.346 [216] | 2.136$e$-5 |
| *VIMSSQ* | 0.308 | 1.911 [216] | 0.057 |
| *Low jitter : Time* | -7.185 | -1.438 [216] | 0.152 |
| *High jitter : Time* | -3.248 | -0.650 [216] | 0.516 |
| *Low jitter : VIMSSQ* | -0.036 | -0.182 [216] | 0.856 |
| *High jitter : VIMSSQ* | 0.537 | 2.695 [216] | 0.008 |
| *VIMSSQ : Time* | -0.283 | -1.420 [216] | 0.157 |
| *Low jitter : Time : VIMSSQ* | 0.080 | 0.284 [216] | 0.777 |
| *High jitter : Time : VIMSSQ* | -0.381 | -1.352 [216] | 0.178 |

Table 3. Linear mixed-effects model results predicting SSQ Total from time, jitter condition, and VIMSSQ. Random intercept only model with VIMSSQ and time centered prior to analysis. Post-test and baseline are the reference conditions.

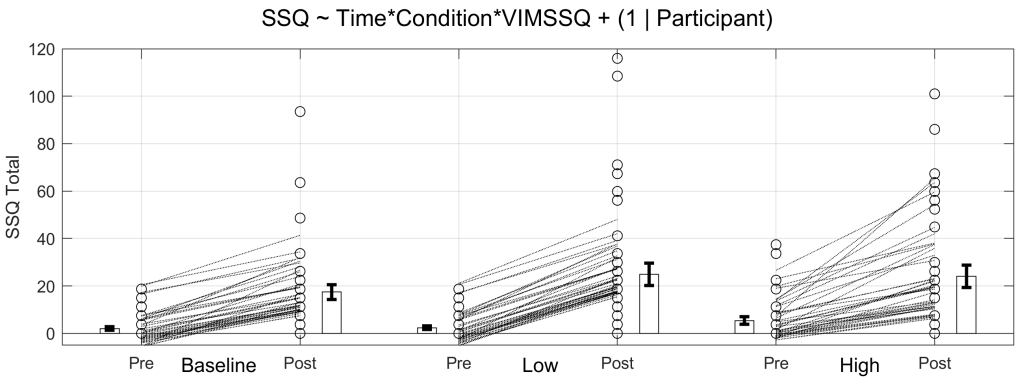

Fig. 12. Linear mixed-effects model results predicting SSQ Total from time, jitter condition, and VIMSSQ (fixed effects). Random intercept only model with VIMSSQ and time centered prior to analysis. White vertical bars depict the mean for each condition (error bars represent +/- 1 SE of the mean).

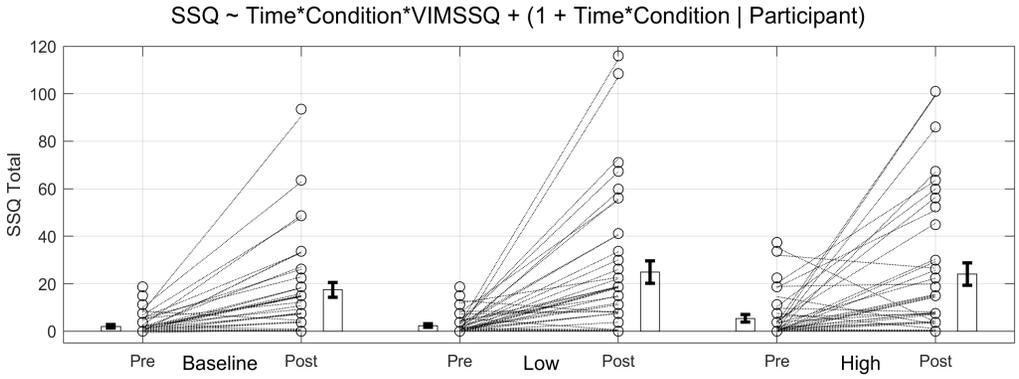

Fig. 13. Linear mixed-effects model results predicting SSQ Total from time and jitter condition. Model includes random slopes for the effects of time and condition (and interactions). White vertical bars depict the mean for each condition (error bars represent +/- 1 SE of the mean).

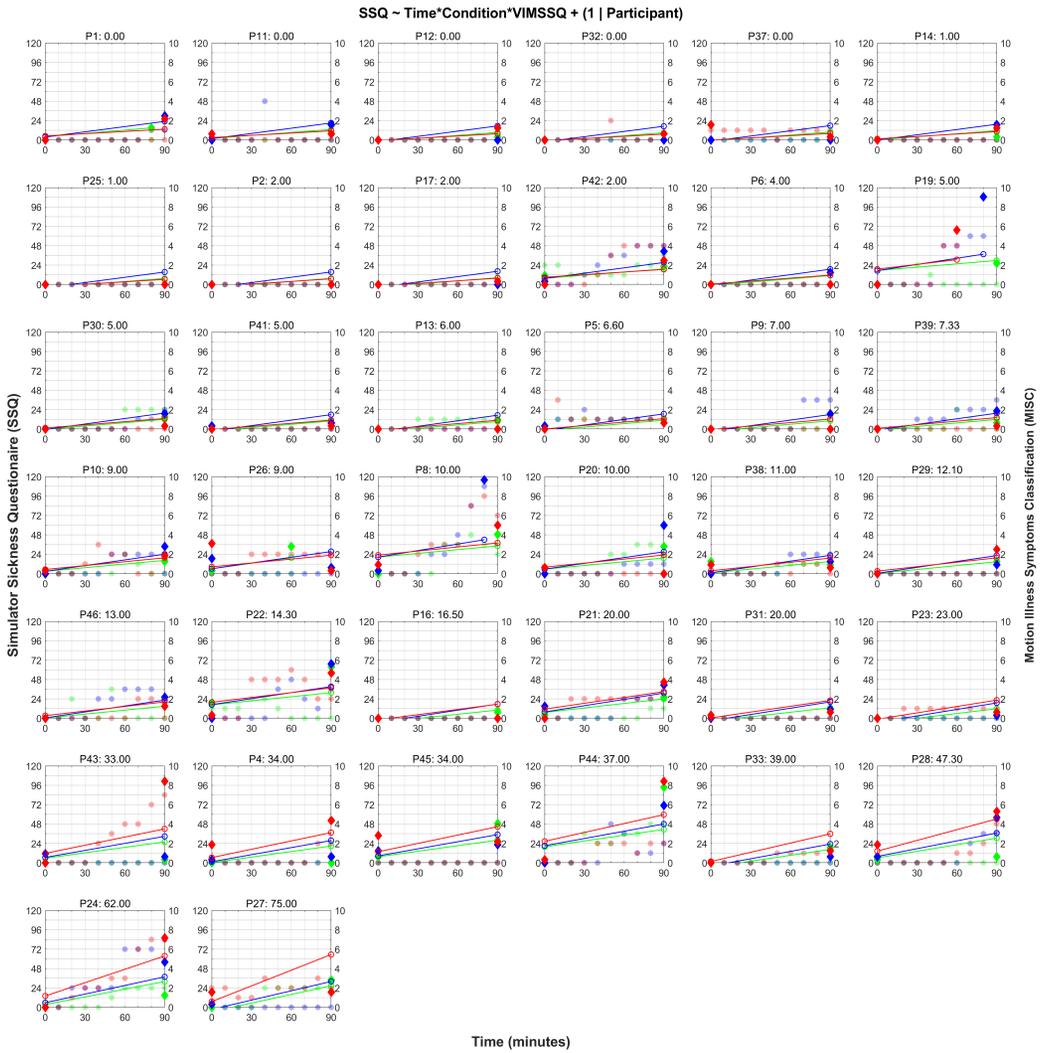

Fig. 14. Individual participant results (sorted by VIMSSQ) for the linear mixed-effects model predicting SSQ Total from time, jitter condition, and VIMSSQ. Model includes fixed effects for time and condition (and interactions).

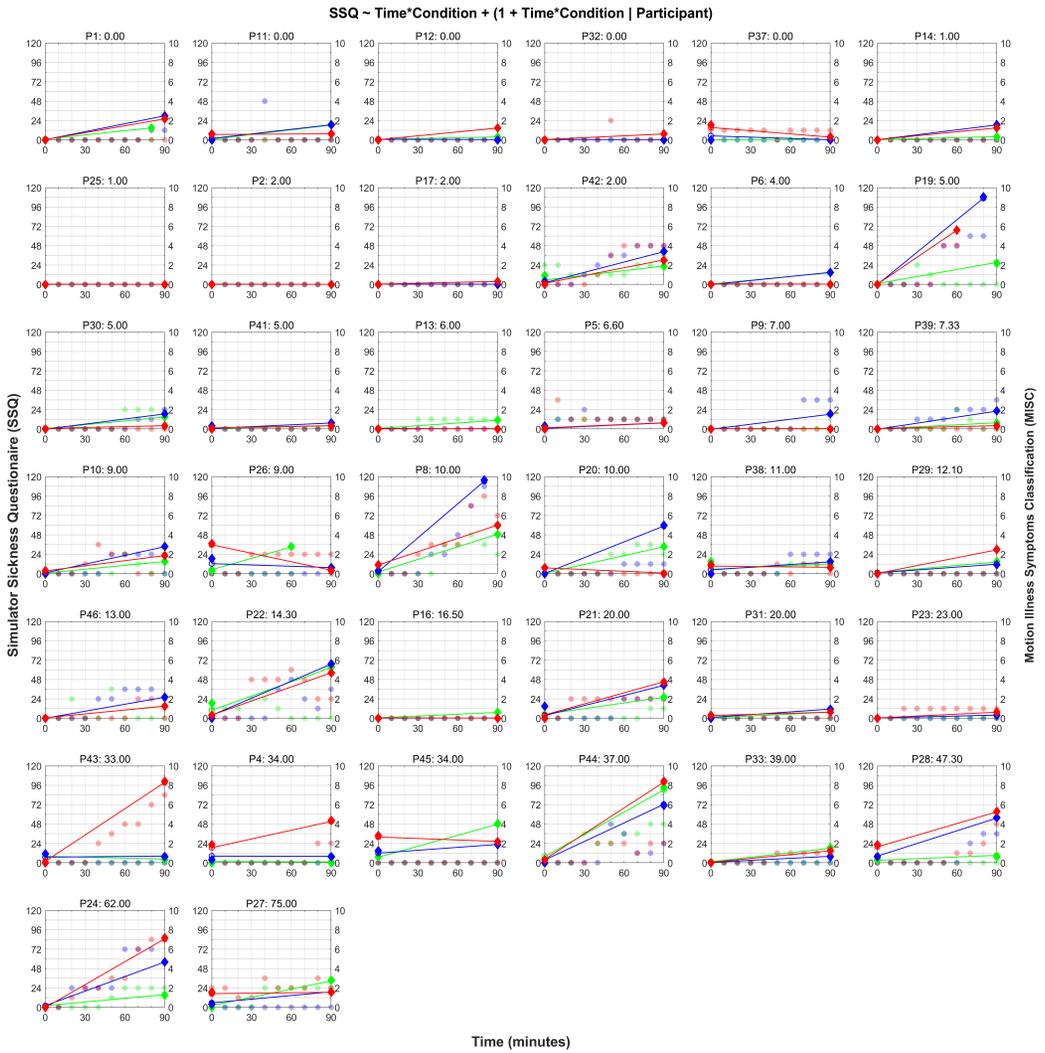

Fig. 15. Individual participant results (sorted by VIMSSQ) for the linear mixed-effects model predicting SSQ Total from time and jitter condition. Model includes random slopes for the effects of time and condition (and interactions).

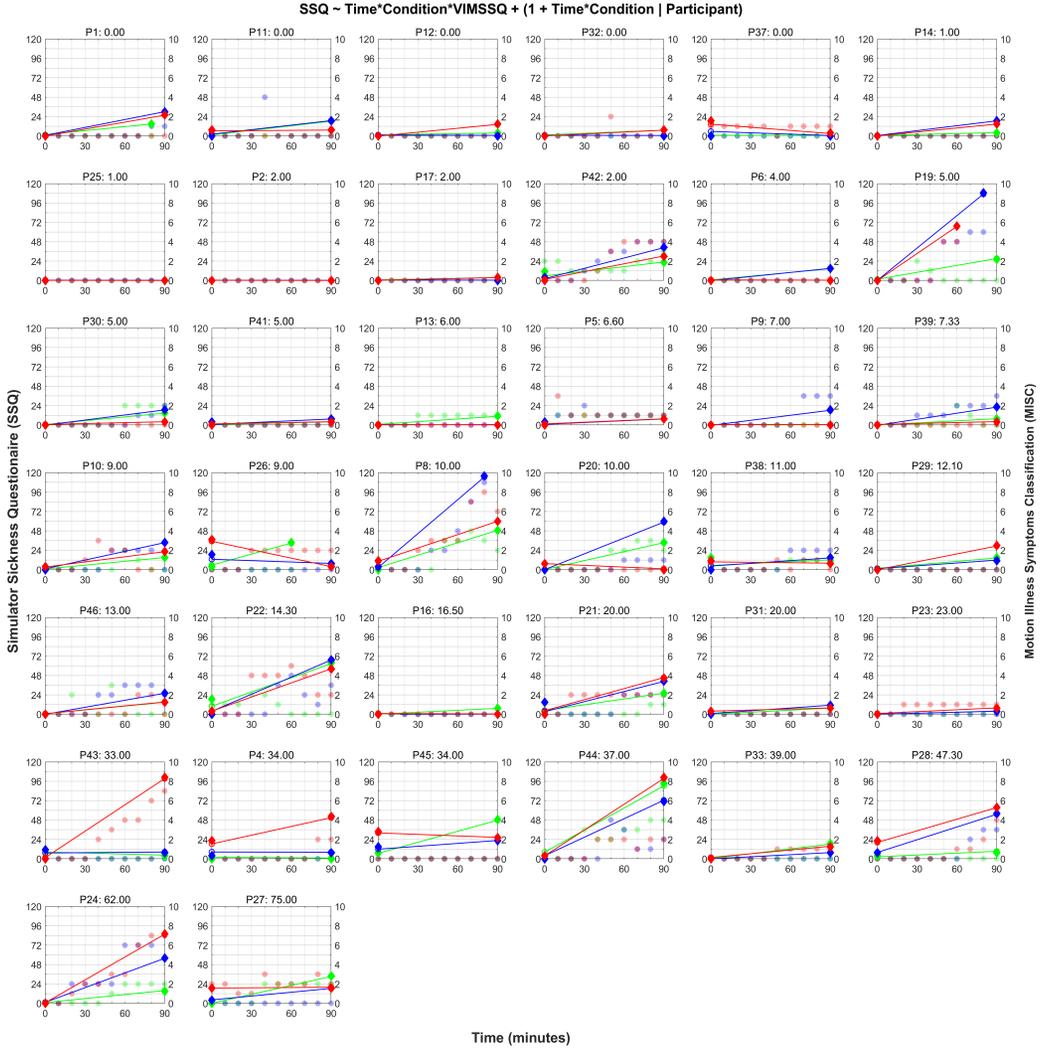

Fig. 16. Individual participant results (sorted by VIMSSQ) for the linear mixed-effects model predicting SSQ Total from time, jitter condition, and VIMSSQ. Model includes random slopes for the effects of time and condition (and interactions).

## 2.4 MISC Models

Identical models to those reported above for SSQ were conducted on MISC scores, but with time now as a continuous predictor (and mean-centered prior to analysis).

In the first (maximal) model assessing only the influence of the combination of jitter condition and time, we fit a mixed-effects linear model of the form:

$$MISC = \beta_{Intercept} + \beta_{Low}X_{1ij} + \beta_{High}X_{2ij} + \beta_{Time}X_{3ij} + \beta_{Low:Time}X_{1:3ij} + \beta_{High:Time}X_{2:3ij} + \beta_{Participant}$$

where $MISC$ is the reported Misery Scale value, $\beta_{Intercept}$ is the predicted MISC score in the baseline condition at t=45 minutes, $\beta_{Low}X_{1ij}$ is the predicted change in MISC at $t$=45 minutes in the low jitter condition (and its random slope), $\beta_{High}X_{2ij}$ is the predicted change in MISC at $t$=45 minutes in the high jitter condition (and its random slope), $\beta_{Time}X_{3ij}$ is the predicted change in MISC for each additional minute in the baseline condition, $\beta_{Low:Time}X_{1:3ij}$ is the predicted additional rate of change in MISC per minute in the low jitter condition (relative to baseline), $\beta_{High:Time}X_{2:3ij}$ is the predicted additional rate of change in MISC per minute in the high jitter condition (relative to baseline), and $\beta_{Participant}$ is a random-effects parameter that fits individual intercepts to each participant to account for between-participants sources of error. The results of this model are reported in Table 4.

| Parameter | Estimate | $t$ value [DoF] | $p$ value |
| --- | --- | --- | --- |
| *Intercept* | 0.361 | 4.074 [1125] | 4.950$e$-5 |
| *Low jitter* | 0.232 | 1.987 [1125] | 0.047 |
| *High jitter* | 0.377 | 2.759 [1125] | 0.006 |
| *Time* | 0.007 | 2.947 [1125] | 0.003 |
| *Low jitter : Time* | 0.010 | 2.576 [1125] | 0.010 |
| *High jitter : Time* | 0.010 | 2.193 [1125] | 0.029 |

Table 4. Linear mixed-effects model results predicting MISC from time and jitter condition. Maximal model with inclusion of random slopes for each effect. Baseline condition is reference group, and time is mean-centered (reference: $t$=45).

In the second (random intercept only) model we included the effect of VIMSSQ (which was mean centered) and its interactions. The results of this model are reported in Table 5.

| Parameter | Estimate | $t$ value [DoF] | $p$ value |
| --- | --- | --- | --- |
| *Intercept* | 0.365 | 3.406 [1119] | 6.822$e$-4 |
| *Low jitter* | 0.205 | 3.154 [1119] | 0.002 |
| *High jitter* | 0.349 | 5.356 [1119] | 1.035$e$-7 |
| *Time* | 0.007 | 4.290 [1119] | 1.946$e$-5 |
| *VIMSSQ* | 0.007 | 1.235 [1119] | 0.217 |
| *Low jitter : Time* | 0.009 | 3.947 [1119] | 8.402$e$-5 |
| *High jitter : Time* | 0.008 | 3.562 [1119] | 3.832$e$-4 |
| *Low jitter : VIMSSQ* | 1.665$e$-5 | 0.005 [1119] | 0.996 |
| *High jitter : VIMSSQ* | 0.017 | 4.563 [1119] | 5.613$e$-6 |
| *VIMSSQ : Time* | 2.552$e$-4 | 2.823 [1119] | 0.005 |
| *Low jitter : Time : VIMSSQ* | -4.001$e$-5 | -0.313 [1119] | 0.754 |
| *High jitter : Time : VIMSSQ* | 3.013$e$-4 | 2.358 [1119] | 0.019 |

Table 5. Linear mixed-effects model results predicting MISC from time, jitter condition, and VIMSSQ. Random intercept only model with VIMSSQ and time mean-centered prior to analysis. Baseline condition is reference group.

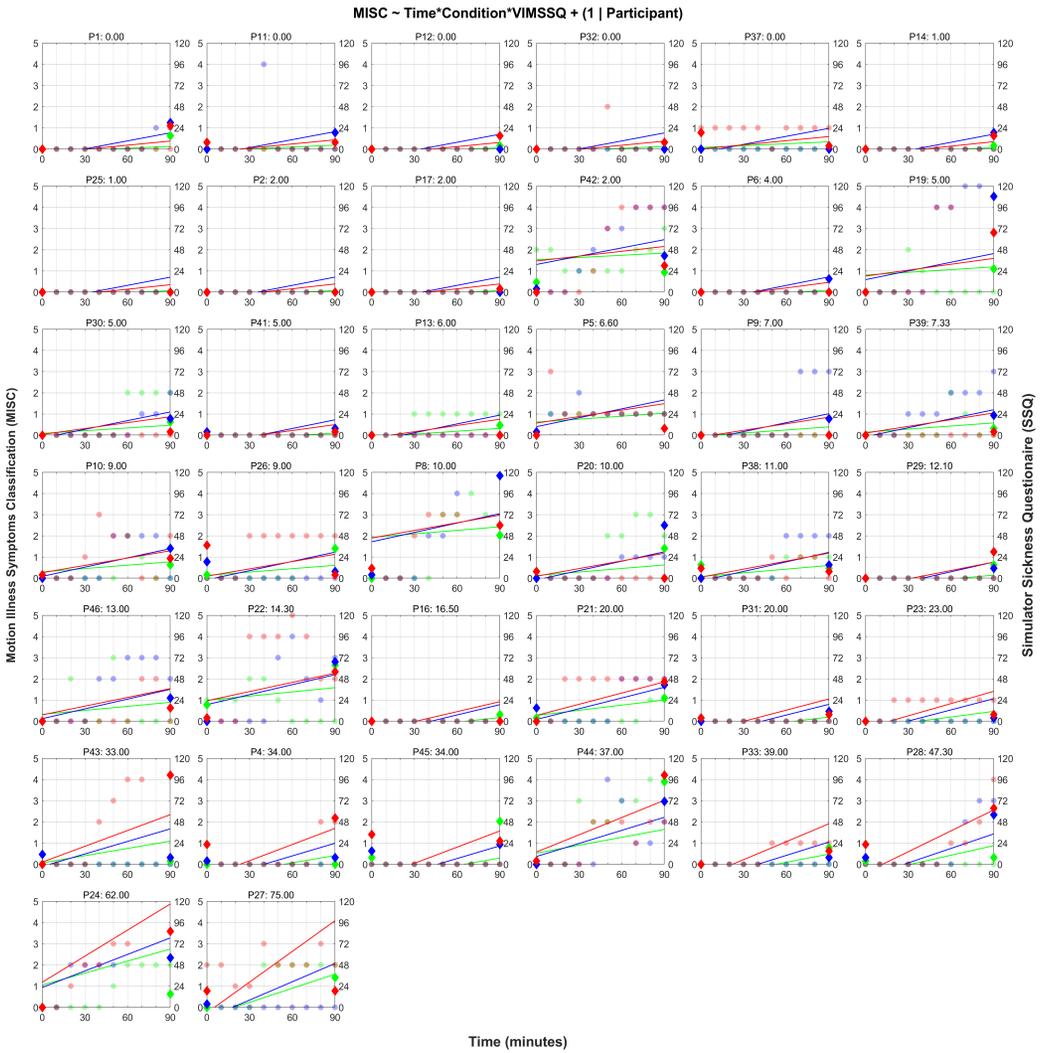

Fig. 17. Individual participant results (sorted by VIMSSQ) for the linear mixed-effects model predicting MISC from time, jitter condition, and VIMSSQ. Model includes fixed effects for time and condition (and interactions). Diamonds on the left and right vertical axes of each plot depict mean SSQ Total (pre and post).

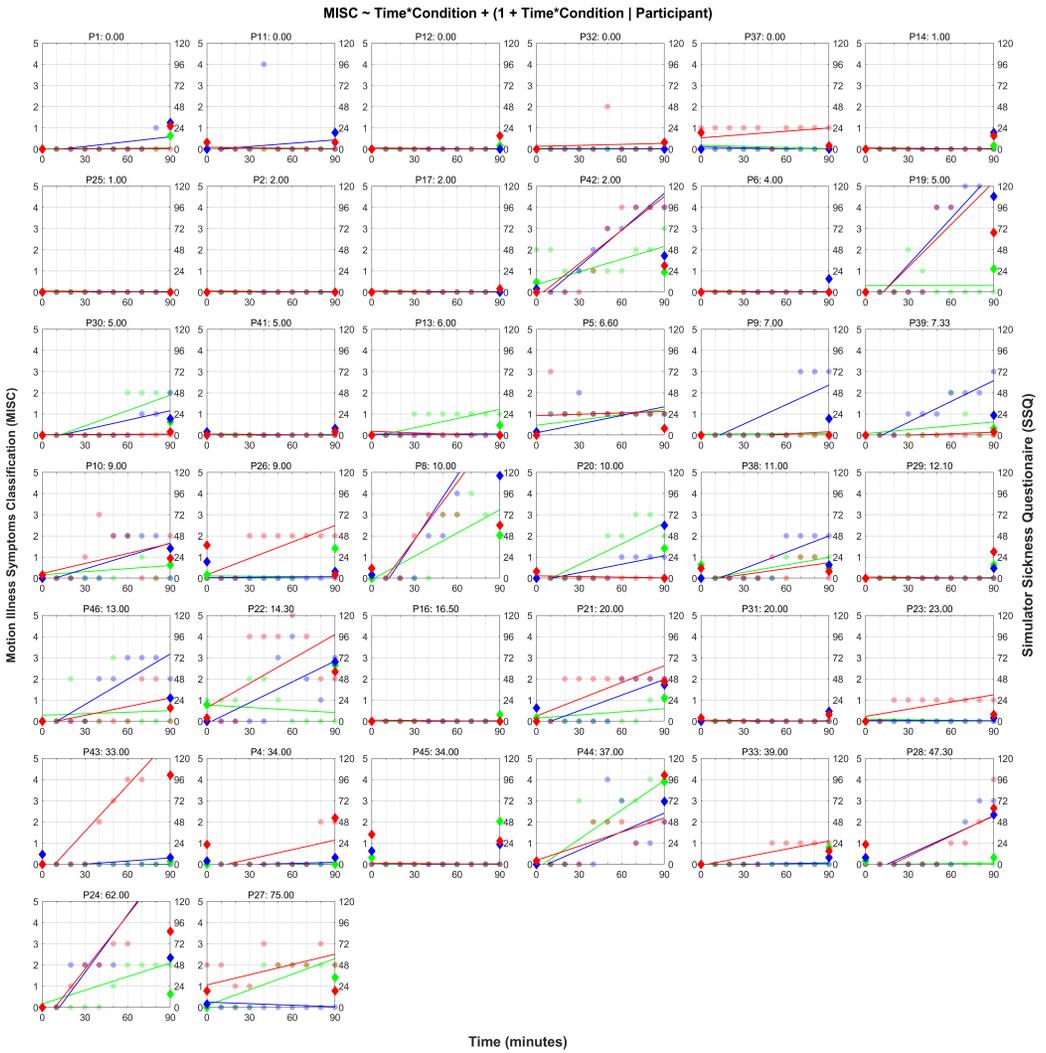

Fig. 18. Individual participant results (sorted by VIMSSQ) for the linear mixed-effects model predicting MISC from time and jitter condition. Model includes random effects for time and condition (and interactions). Diamonds on the left and right vertical axes of each plot depict mean SSQ Total (pre and post).

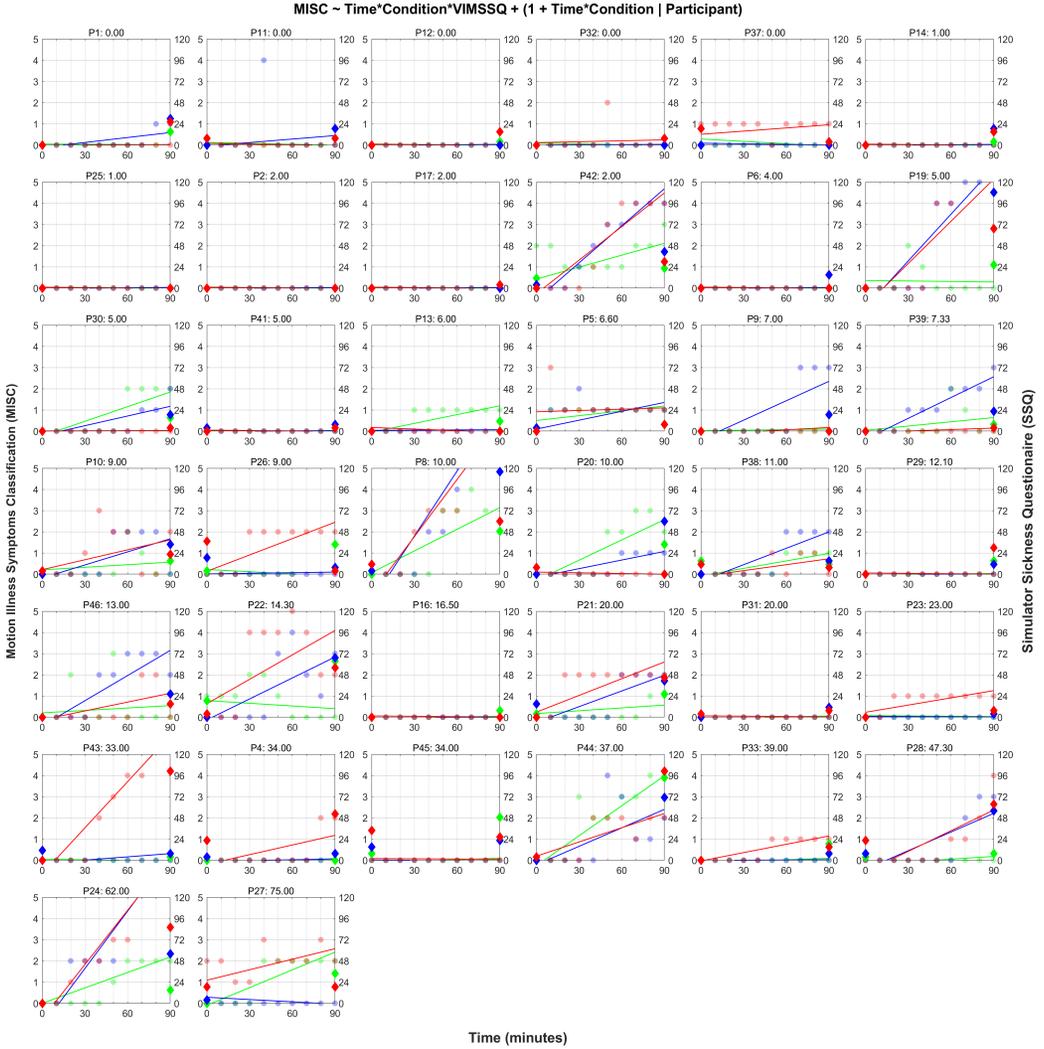

Fig. 19. Individual participant results (sorted by VIMSSQ) for the linear mixed-effects model predicting MISC from time, jitter condition, and VIMSSQ. Model includes random effects for time and condition (and interactions). Diamonds on the left and right vertical axes of each plot depict mean SSQ Total (pre and post).

## 2.5 Eye Strain

Identical models to those reported above for MISC were conducted on eye strain scores. In the first (maximal) model assessing only the influence of the combination of jitter condition and time, we fit a mixed-effects linear model of the form:

$$Eye\ Strain = \beta_{Intercept} + \beta_{Low}X_{1ij} + \beta_{High}X_{2ij} + \beta_{Time}X_{3ij} + \beta_{Low:Time}X_{1:3ij} + \beta_{High:Time}X_{2:3ij} + \beta_{Participant}$$

where $EyeStrain$ is the reported eye strain value, $\beta_{Intercept}$ is the predicted eye strain score in the baseline condition at t=45 minutes, $\beta_{Low}X_{1ij}$ is the predicted change in eye strain at $t$=45 minutes in the low jitter condition (and its random slope), $\beta_{High}X_{2ij}$ is the predicted change in eye strain at $t$=45 minutes in the high jitter condition (and its random slope), $\beta_{Time}X_{3ij}$ is the predicted change in eye strain for each additional minute in the baseline condition, $\beta_{Low:Time}X_{1:3ij}$ is the predicted additional rate of change in eye strain per minute in the low jitter condition (relative to baseline), $\beta_{High:Time}X_{2:3ij}$ is the predicted additional rate of change in eye strain per minute in the high jitter condition (relative to baseline), and $\beta_{Participant}$ is a random-effects parameter that fits individual intercepts to each participant to account for between-participants sources of error. The results of this model are reported in Table 6.

| Parameter | Estimate | $t$ value [DoF] | $p$ value |
| --- | --- | --- | --- |
| *Intercept* | 1.539 | 15.922 [1125] | 1.250$e$-51 |
| *Low jitter* | 0.085 | 1.010 [1125] | 0.313 |
| *High jitter* | 0.056 | 0.628 [1125] | 0.530 |
| *Time* | 0.010 | 5.021 [1125] | 5.977$e$-7 |
| *Low jitter : Time* | 0.003 | 1.467 [1125] | 0.143 |
| *High jitter : Time* | 0.002 | 0.696 [1125] | 0.487 |

Table 6. Linear mixed-effects model results predicting eye strain from time and jitter condition. Maximal model with inclusion of random slopes for each effect. Baseline condition is reference group, and time is mean-centered (reference: $t$=45).

In the second (random intercept only) model we included the effect of VIMSSQ (which was mean centered) and its interactions. The results of this model are reported in Table 7.

| Parameter | Estimate | $t$ value [DoF] | $p$ value |
| --- | --- | --- | --- |
| *Intercept* | 1.543 | 19.557 [1119] | 1.713$e$-73 |
| *Low jitter* | 0.075 | 1.770 [1119] | 0.077 |
| *High jitter* | 0.048 | 1.137 [1119] | 0.256 |
| *Time* | 0.010 | 9.541 [1119] | 8.531$e$-21 |
| *VIMSSQ* | 0.010 | 2.314 [1119] | 0.021 |
| *Low jitter : Time* | 0.002 | 1.640 [1119] | 0.100 |
| *High jitter : Time* | 0.001 | 0.859 [1119] | 0.390 |
| *Low jitter : VIMSSQ* | -0.003 | -1.118 [1119] | 0.264 |
| *High jitter : VIMSSQ* | 0.006 | 2.492 [1119] | 0.013 |
| *VIMSSQ : Time* | -4.406$e$-6 | -0.075 [1119] | 0.940 |
| *Low jitter : Time : VIMSSQ* | 1.152$e$-4 | 1.391 [1119] | 0.164 |
| *High jitter : Time : VIMSSQ* | 2.864$e$-4 | 3.458 [1119] | 5.641$e$-4 |

Table 7. Linear mixed-effects model results predicting eye strain from time, jitter condition, and VIMSSQ. Random intercept only model with VIMSSQ and time mean-centered prior to analysis. Baseline condition is reference group.